
\def\bigtype{\let\rm=cmr12  \let\it=cmti12 \let\bf=cmbx12
\let\sl=cmsl12
 \baselineskip=18pt }

\def\d#1/d#2{ {\partial #1\over\partial #2} }

\newcount\probno

\def\prob{\global\advance\probno by1 \the\probno }

\newcount\itemno
\def\item{  \global\advance\itemno by 1
(\romannumeral\the\itemno)
}

\newcount\probno
\def\prob{\global\advance\probno by1 \the\probno
\global\itemno=0}

\newcount\probsetno
\def\probset{\global\advance\probsetno by1
\centerline{Problem Set \the\probsetno } }

\newcount\topicno
\def\topic{\global\advance\topicno by1 \the\topicno }

\newcount\partno
\def\part{ PART\ \global\advance\partno by 1
\uppercase\expandafter{\romannumeral\the\partno}.
\global\probno=0\global\lectno=0
}

\newcount\lectno
\def\lect{ \centerline{LECT\ \global\advance\topicno by1
\the\lectno.} }

 \newcount\sectno
\def\sect #1 { {\it \global\advance\sectno by1 \the\sectno . #1} }

\def\pdr{\partial}

\def\eps{\epsilon}
\def\half{{1\over 2}}
\def\tr{\hbox{tr}}

\def\mod{\;\;{\rm mod}\;\;}

\def\linebreak{\hfil\break}

\def\rochester{
 \centerline{\it Department of Physics and Astronomy}
 \centerline{\it University of Rochester}
 \centerline{\it Rochester, N.Y. 14627}
               }


\newcount\eqnumber
\def\beq{ \global\advance\eqnumber by 1 $$ }
\def\eeq{ \eqno(\the\eqnumber)$$ }
\def\n{\global\advance \eqnumber by 1\eqno(\the\eqnumber)}
\def\puteqno{
\global\advance \eqnumber by 1 (\the\eqnumber)}
\def\beqs{$$\eqalign}
\def\eeqs{$$}


\def\ifundefined#1{\expandafter\ifx\csname
#1\endcsname\relax}

\newcount\refno \refno=0  
\def\[#1]{
\ifundefined{#1}
\advance\refno by 1
\expandafter\edef\csname #1\endcsname{\the\refno}
\fi[\csname
#1\endcsname]}
\def\refis#1{\noindent\csname #1\endcsname. }

\def\label#1{
\ifundefined{#1}
\expandafter\edef\csname #1\endcsname{\the\eqnumber}
\else\message{label #1 already in use}
\fi{}}
\def\(#1){(\csname #1\endcsname)}
\def\eqn#1{(\csname #1\endcsname)}

\baselineskip=15pt
\parskip=10pt

\magnification=1200

\def\BEGINIGNORE#1ENDIGNORE{}

\baselineskip=20pt

\def\un#1{\underline{#1}}


 \def\for{ \;\;{\rm for }\;\;}

\magnification=1200
UR 1256

ER 40685-708

hepth@xxx/9204076
\vskip.5in
\centerline{\bf Baryons as Solitons in Three Dimensional Quantum
Chromodynamics}
\vskip.5in

\centerline{ G. Ferretti,  S.G.Rajeev and Z. Yang}

\rochester
\vskip.5in

\centerline{\bf Abstract} We show that baryons of three dimensional
Quantum Chromodynamics can be understood as solitons of its  effective
lagrangian. In the parity preserving phase we study, these baryons are fermions
for odd $N_c$ and bosons for even  $N_c$, never anyons. We quantize the
collective variables of the solitons and there by calculate the flavor quantum
numbers, magnetic moments and mass splittings of the baryon. The flavor quantum
numbers are in agreement with naive quark model for the low lying states. The
magnetic moments and mass splittings are smaller in the soliton model by a
factor of $\log {F_\pi\over N_c m_\pi}$. We also show that there is a dibaryon
solution that is an analogue of the deuteron. These solitons can describe
defects in a quantum anti--ferromagnet.

\vfill\break

\sect{ Introduction}

In accompanying paper\[I], we constructed  the low energy effective lagrangian
of the `mesons' of  three dimensional QCD,  with $N_c$ colors and $2n$
flavors.It is a nonlinear sigma model with the Grassmannian
$Gr_{n}=SU(2n)/S(U(n)\times U(n))$  as the target space.  The field variables
$\chi$  and $A$ take values in the group $SU(2n)$ and the Lie algebra
$\un{G}=\un{SU(n)}\oplus \un{SU(n)}\oplus R$ respectively. The effective action
is  \beqs{ S&=S_1+S_k+S_m\cr &={{F_\pi}\over
2}\int\tr\nabla_{\mu}\chi^{\dag}\nabla^{\mu}\chi d^3x\cr &+{k\over 4\pi}\int
\tr \left[ A_1dA_1+{2\over 3}A_1^3\right]-{k\over 4\pi}\int \tr \left[
A_2dA_2+{2\over 3}A_2^3\right]\cr & + {{F_\pi}\over 2}m_{\pi}^2\int \tr
\eps\chi\eps\chi^{\dag} d^3x.\cr }\eeqs  Here, $A=A_1+A_2+A_3$  is a gauge
field valued in $\un{SU(n)}\oplus \un{SU(n)}\oplus R$  Lie algebra. The
covariant derivative is  \beq \nabla_\mu\chi=\pdr_\mu\chi-i\chi A_\mu. \eeq
Also, $\eps=\pmatrix{1&0\cr 0&-1\cr}$ is the matrix that commutes with $G$.

The Chern--Simons terms are necessary to realize the discrete symmetries of
3DQCD correctly. Comparison with 3DQCD shows that the level number $k$  of the
Chern-- Simons theory is $N_c$, the number of colors. This lagrangian describes
pseudo--scalars (`pions') of mass $m_\pi$ and vector mesons of mass ${2\pi
F_{\pi}\over k}$.

We will show in this paper that this effective lagrangian also describes the
baryons: they are the    topological soliton solutions.  The ideas are very
similar to those in four dimensions. (For a review, see Ref. \[balreview]). The
Chern--Simons terms provide the  short repulsion  necessary for the stability
of these solitons\[hongetal] \[jackiwetal].  We argue that these solitons are
the baryons of 3DQCD. We then study the low energy properties (mass
splittings, magnetic moments, flavor quantum numbers) by using an effective
lagrangian for the collective motion. This effective lagrangian  (for the case
of four flavors, $n=2$), is a  $0+1$ dimensional nonlinear sigma model on the
coset space  $S(U(2)\times U(2))/U(1)\times U(1)$. We find that the quantum
numbers of the solitons are the same,  (at low energies) as those of the
baryons in the quark model. However, the mass splittings are smaller than in
the naive quark model. This is because the size of the soliton is bigger than
that of the baryon in the quark model.

A summary of  the behaviour of our effective action under discrete symmetries
is  perhaps useful. There are three discrete transformations $P_0,\sigma$ and
$P_2$ of interest. \beqs{ P_0:\chi(x_1,x_2,t)\to \chi(-x_1,x_2,t)&\quad
A_{1,2,3}(x,t)\to A_{1,2,3}(-x_1,x_2,t)\cr \sigma:\chi(x,t)\to
\chi(x,t)\pmatrix{0&1\cr 1&0\cr}&\quad A_1(x,t)\leftrightarrow A_2(x,t)\quad
A_3(x,t)\to -A_3(x,t)\cr P_2:\chi(x,t)\to \pmatrix{0&1\cr
1&0\cr}\chi(x,t)&\quad A_{1,2,3}(x,t)\to A_{1,2,3}(x,t).\cr }\eeqs Under $P_0$,
the terms $S_1$ and $S_m$ are invariant; under $\sigma$, only $S_1$  is
invariant. Also, $P_2$ leaves $S_1$ and $S_k$ invariant. Thus the only symmetry
of the total effective action is the product $P_0\sigma P_2=P_1P_2=P$. It is
this product that we should identify with physical parity.

\sect { Baryons of 3DQCD as Solitons }

As in the four dimensional Skyrme model, we should expect baryons to arise as
solitons of this effective lagrangian. In fact we will see now that there are
such solitons. Furthermore we will show that they are fermions when $k$ is odd
(bosons when $k$ is even) and that their wavefunctions transform under the
flavor symmetry $G$ as expected from the quark model. We will consider only the
special case $n=2$ in detail. (This is analogous to the Skyrme model with
$SU(3)$ symmetry \[balreview].) The cases $n>2$ are  exactly analogous and
nothing much is learned by being more general. The case $n=1$ has some special
features because the third homotopy group of the coset space (which is $CP^1$)
is also non--trivial; we will comment on it later.

The Grassmannian has nontrivial second homotopy group, which allows for the
existence of topological solitons. The topological current is, in our choice of
variables,  \beq j_{\mu}={1\over 2\pi}\eps_{\mu\nu\rho}\pdr^{\nu}A_3^{\rho}
.\eeq  The  soliton number is just the vorticity of $A_3$ at infinity: \beq
Q={1\over 2\pi}\int d\theta A_{3\theta}. \eeq This current is to be identified
with baryon number current of 3DQCD. For, it is equal to the expectation value
${1\over N_c}<\sum_i \bar q^i\gamma_\mu q_i>$  coupled to an external $A_3$
gauge field. (To be precise, this vacuum expectation value is equal to the
above topological current up terms   that do not contribute to the total
charge)\[goldwilcz],\[baletal] \[witten].

A rotationally symmetric ansatz for the static solution is  \beq
\chi_0(r,\theta)=\pmatrix{ 1&0&0&0\cr  0&\cos\phi(r) e^{i{Q\theta}}&
\sin\phi(r) &0\cr 0&-\sin\phi(r) &\cos\phi(r)e^{-i{Q\theta}}&0\cr 0&0&0&1\cr}
\eeq and   \beqs{ A_{\mu}^0dx^{\mu}&=\half V(r)dt \pmatrix{1&0&0&0\cr
0&-1&0&0\cr 0&0&-1&0\cr 0&0&0&1} -\half A(r)d\theta\pmatrix{1&0&0&0\cr
0&-1&0&0\cr 0&0&1&0\cr 0&0&0&-1\cr}\cr &+\half [\tilde V(r) dt +\tilde
A(r)d\theta]\pmatrix{1&0&0&0\cr 0&1&0&0\cr 0&0&-1&0\cr 0&0&0&-1\cr }\cr }\eeqs
$V,\tilde V, \tilde A$ are Lagrange multipliers which can be eliminated by
their equations of motion: \beq V(r)= {k\over 2\pi F_{\pi}r}A'(r), \quad \tilde
V=0, \quad \tilde A(r)= Q\cos^2\phi(r) .\eeq

In the above  ansatz, $Q$ is the soliton number of the solution. (We will
first study $Q=1$, but later we will need $Q=2$ for the dibaryon.).  The
boundary conditions on $\phi$ and $A$ are \beq \phi(0)={\pi\over
2},A(0)=0\quad\hbox{and}\quad\phi(r)\to 0,A(r)\to Q \quad\hbox{as}\quad
r\to\infty \eeq After substituting into the lagrangian and eliminating
$V,\tilde V, \tilde A$, we get the energy integral \beqs{ E(\phi,A)&=\pi
F_{\pi}\int_0^\infty[2r\phi'^2+{A^2\sin^2\phi+(Q-A)^2\cos^2\phi +{Q^2\over
4}\sin^2 2\phi \over r}\cr &+\left({k\over 2\pi F_\pi}\right)^2{A'^2\over r}
+4m_{\pi}^2r \sin^2\phi ]dr \cr }\eeqs It is clear that under scaling \beq
\phi(r)\to \phi(\lambda r)\quad A(r)\to A(\lambda r)  \eeq the first two terms
are invariant, the third goes like $\lambda^{-2}$ and the last  term like
$\lambda^2$. This shows that the solution is stable under scaling ("Derrick's
Theorem").  The mass term tends to shrink the soliton while the vector mesons
provide a short range repulsive force which tends to expand it.

The Euler--Lagrange equations for the cases $Q=1,2$ can be solved numerically,
by relaxation methods. In Fig. 1 we plot the solution for $Q=1$. The energy of
the baryon (Fig. 2) is almost a linear function
$E\sim a {{2\pi F_\pi}\over {N_c}}+ b m_{\pi}$,
with a numerical fit to the constants, $a\sim 1.058, b\sim 1.167$.  This
almost linear dependence can be understood by a variational argument (an
approximate `virial theorem').
We will comment on the $Q=2$  solution later.

In the limit $m\neq 0, k=0$, the soliton
will shrink to a point.In the other  limit
$m\to 0$ keeping  $k$ fixed, it will expand to infinite size.  It is a
surprising fact that, in the absence of a current quark mass, the soliton will
expand to infinity and disappear. In general, the size of the soliton is of the
same order as the pion Compton wavelength, which is very different from the
situation in four dimensions.

If both $m$ and $k$ are zero the solution tends to the well--known soliton of
the $CP^1$ model, \beq \phi={\pi\over 2}-\arctan{r\over a}, \quad
A(r)={{r^2}\over{a^2+r^2}} .\eeq This case is scale invariant, so that there is
a soliton of every possible size $a$. However, in this limit, (which is not
related to 3DQCD since $k=0$)  the moment of inertia of the soliton is
infinite\[Al], leading to a spontaneous breaking of rotation invariance.

One can argue on purely topological grounds that these solitons are fermions
when $N_c$ is odd and bosons when $N_c$ is even. First one shows that they are
of spin ${N_c\over 2} {\rm mod} Z$ by following an argument analogous to Witten
\[witten]. One consider a closed path in the configuration space which
corresponds to creating a soliton anti--soliton pair, separating them to a
large distance, then rotating the soliton through $2\pi$ and then annihilating
them. This process has a probability amplitude $(-1)^{N_c}$. (We don't give the
details  since the following collective co--ordinate method will show quite
explicitly that the spin is $N_c/2 {\rm mod} Z$).
Then one can use the general spin
statistics theorems\[spinstatistics] of soliton theories to show that they are
fermions (bosons) for odd $N_c$ (even $N_c$).

\sect { Collective Coordinate  Quantization}

The three dimensional sigma model has a global invariance under $G=S(U(n)\times
U(n))$. Therefore given any static solution $\chi_0(x), $ we can find another
one, $X \chi_0(x)$ for $X\in G$, of the same energy. By allowing $X$ to be a
slowly varying function of time, we will excite  the lowest energy states  of
the soliton. However, not all such rotations produce a physically distinct
soliton: there is  a subgroup $H$ of $G$  that changes $\chi_0$ only by a gauge
transformation. Thus the configuration space of the collective motion of the
soliton is a coset space $G/H$. In fact, the effective action for  collective
motion is a {\it one dimensional } nonlinear sigma model on $G/H$. This can be
described again by a variable $X$ valued in $G$ and a one dimensional gauge
field  valued in $\un{H}$. We will see that the Chern--Simons term of the three
dimensional sigma model induces a  Chern--Simons term for the one dimensional
theory as  well. This term will then dominate  the low energy properties of the
soliton.(In particular, how the soliton wave function transforms under $G$).

Let us first determine $H$, the subgroup of $G$ that only changes $\chi_0$ by a
gauge transformation. Recall that the gauge transformations of the three
dimensional theory are right multiplications by $G$ while the global symmetry
is  a left multiplication. Thus $h$ is in $H$ if $\chi_0^{\dag}(x) h \chi_0(x)$
is in $G$ for all $x$. A short calculation will show that such elements are of
the form  \beq h=\pmatrix{h_1&0&0&0&\cr 0&e^{i\alpha}&0&0\cr
0&0&e^{i\alpha}&0\cr 0&0&0&h_2\cr} \eeq  where $h_{1,2}$ are in $U(n-1)$. (It
is useful to consider first the special case $r=0$ which will already require
that $h$ be block diagonal.) The dimension of $G/H$ is then  $4n-3$. (There are
also two translational collective modes which we are ignoring. They can be
taken care of trivially \[rajaraman]. Our soliton has  one less degree of
freedom than the instanton of the two dimensional Grassmannian sigma
model\[perelemov], because scale invariance is not symmetry. In the limit
$m=k=0$ we recover this collective mode).

For a more detailed study, we will restrict to the case $n=2$.Then $H$ is the
abelian group $U(1)\times U(1)$, generated by  \beq  y=\pmatrix{1&0&0&0\cr
0&-1&0&0\cr 0&0&-1&0\cr 0&0&0&1\cr}\quad  \;\;\hbox{\rm and }\;\; \tilde
y=\pmatrix{1&0&0&0\cr 0&0&0&0\cr 0&0&0&0\cr 0&0&0&-1\cr}. \eeq The collective
variables $X,a,\tilde a$ describe a deformation of the soliton configuration,
\beq \chi(x,t)=X(t)\chi_0(x) \eeq and  \beq A(x,t)=
A_{\mu}^0dx^{\mu}+[a(t)y+\tilde a(t) \tilde y]dt
-iP_{\un{G}}\left(\chi_0^{\dag}X^{\dag}\dot X\chi_0\right)dt, \eeq where $A^0$
is the static solution.

The collective action will be a nonlinear sigma model on $G/H$ with $a,\tilde
a$ playing the role of one dimensional gauge fields valued  in $H$.

The  general one dimensional nonlinear sigma model on $G/H$ has action  \beqs{
S_{\rm coll}&=\half  I_1 \int \tr D_tX_1^{\dag}D_tX_1 dt+ \half I_2\int  \tr
D_tX_2^{\dag}D_tX_2 dt+\cr &    \half  I_3\int (\dot \xi+ \tilde a)^2 dt
+\mu\int a dt+ \tilde \mu \int \tilde a dt\cr }\eeqs up to higher derivatives
in
$t$. Here, \beq X=\pmatrix{e^{i\xi\over 2}X_1&0\cr 0&e^{-i{\xi\over 2}}
X_2\cr}, \for X_{1,2}\in SU(2) \eeq and  \beqs{ D_tX_1&=\dot
X_1+X_1(ia\tau_3+ {1\over 2}
i\tilde a\tau_3)\cr D_tX_2&=\dot X_2+X_2(-ia\tau_3+{1\over 2}i\tilde
a\tau_3)\cr
D_{t}\xi=\dot \xi + \tilde a.
\cr }\eeqs The $I_{1,2,3}$ are `moments of inertia' determined by the
microscopic theory.

This action has a gauge invariance (up to boundary terms) under the right
action
of $H$, \beqs{ X_{1}&\to X_{1}e^{i(\lambda+\half\tilde\lambda)\tau_3},\cr
X_{2}&\to X_{2}e^{i(-\lambda+\half\tilde\lambda)\tau_3},\cr \xi&\to \xi+\lambda
\cr a\to a-\dot\lambda,\;\;\; &\tilde a \to  \tilde a -\dot {\tilde \lambda}.
}\eeqs The Chern--Simons terms of the one dimensional gauge  fields  are
linear. In order for $e^{iS_{\rm coll}}$ to  be gauge invariant, $\mu,\tilde
\mu$ have to be integers. Of course, there is no curvature for such a gauge
field. The Chern--Simons term will contribute a phase to the wave function
which will determine its transformation properties  under the global symmetry.

Parity invariance imposes certain relations among the constants of the
effective lagrangian.  Under parity, \beq P:\chi(r,\theta,t)\to \sigma
\chi(r,-\theta, t)\sigma. \eeq One can check that  \beq \sigma
\chi_0(r,-\theta)\sigma= \tau_2\chi_0(r,\theta) \tau_2 \eeq where \beq
\tau_2=\pmatrix{0&-i&0&0\cr i&0&0&0\cr 0&0&0&-i\cr 0&0&i&0\cr}. \eeq Then, we
have  \beq P: X\chi_0(r,\theta)\to \sigma X(t)\sigma  \tau_2 \chi_0(r,\theta)
\tau_2. \eeq The last factor is an element of $G$, so is a gauge
transformation. Hence parity acts on the collective variable as follows: \beq
P:X\to \sigma X\sigma \tau_2. \eeq

Therefore, in the collective action, \beq P:X_1\leftrightarrow X_2
\pmatrix{0&-i\cr i&0\cr},\quad \xi \to -\xi, \quad
a\to a,\quad \tilde a\to -\tilde a \eeq so that
we get  \beq I_1=I_2\quad \tilde \mu=0. \eeq If it weren't for the
Chern--Simons terms, there would have been an  additional discrete symmetry,
\beq X_1\leftrightarrow X_2 \qquad a\to -a\quad  \tilde a\to \tilde a. \eeq
Thus by arguments entirely analogous to the ones that led to the three
dimensional effective action, we get  \beqs{ S_{\rm coll}&=\half I_1 \int \tr
D_tX_1^{\dag}D_tX_1 dt+ \half I_1 \int \tr D_tX_2^{\dag}D_tX_2 dt+\cr & \half
I_3\int (\dot \xi+ \tilde a)^2 dt  +\mu\int a dt  \cr }\eeqs

Now we can determine the constants $I_{1,3}$ and $\mu$ from our `microscopic'
theory, which is our three dimensional effective action. For $I_{1,3}$ we get
the integrals\footnote{\ddag}{ These will diverge in the limit $m=k=0$. This is
the problem noted in \[Al]. It appears to be coincidence that $I_1=I_3$.} \beq
I_1=I_3=\pi \int_0^{\infty} \sin^2 2\phi\;\; r dr \eeq

To calculate   the Chern--Simons term we will need the projections  \beqs{
A_1&= \left[(a+\half \tilde a-\half V) dt -\half A d\theta\right]
\pmatrix{1&0\cr 0&-1}\cr \quad A_2&=\left[(-a+\half \tilde a-\half V) dt -\half
A d\theta\right] \pmatrix{1&0\cr 0&-1}.\cr }\eeqs Since $A$ lies in an abelian
subgroup of $G$, the $A^3$ terms in the Chern--Simons term  will not
contribute. Also, we can drop the terms of order zero in $a, \tilde a$ since
they are part of  the energy of the static soliton.  Thus we  get  \beq \int
\tr A_1dA_1  -\int \tr A_2dA_2=  2\int a dt  \int dr d\theta A'=4\pi Q\int a
\;dt \eeq since $\int dr d\theta  {\pdr A\over \pdr r}=2\pi Q$.($Q$ is  the
soliton   number). Thus we see that $\mu=kQ,\tilde \mu=0$ and, the collective
action is  \beqs{ S_{\rm coll}&=\half I_1 \int \tr D_tX_1^{\dag}D_tX_1 dt+
\half I_1 \int \tr D_tX_2^{\dag}D_tX_2 dt+\cr &    \half I_3\int (\dot
\xi+\tilde a)^2 dt  +kQ\int a dt  \cr }\eeqs

Let us now consider the wave functions of the soliton. They can be thought of
as functions of $X$ that satisfy a constraint coming from the $H$ gauge
invariance. The wave function of the soliton would have  been invariant under
the right action of $H$ if  this collective action were gauge invariant. But,
$S_{\rm coll}$ is invariant only up to a boundary term so that the wave
functions are invariant only up to a phase: \beq \psi(Xe^{(i\lambda y+i\tilde
\lambda \tilde y)})=e^{-ikQ\lambda}\psi(X) .\eeq They are sections of a
nontrivial line bundle on $G/H$.

Left multiplication by $G$ will leave this constraint invariant, so that the
states can be classified by representations of this flavor symmetry.
Furthermore, the  right multiplication  \beq X\to X \exp\left[i{\alpha\over 2}
\pmatrix{0&0&0&0\cr 0&1&0&0\cr 0&0&-1&0\cr 0&0&0&0\cr}\right] \eeq will leave
the constraint invariant. This  symmetry, corresponds to spatial rotations of
the soliton.  For,under spatial rotations,  \beqs{ X(t)\chi_0(r,\theta)\to
X(t)\chi_0(r,\theta+\alpha)&=\cr X(t) \exp\left[i{\alpha\over 2}
\pmatrix{0&0&0&0\cr 0&1&0&0\cr 0&0&-1&0\cr 0&0&0&0\cr}\right]&\chi_0(r,\theta)
\exp\left[i{\alpha\over 2} \pmatrix{0&0&0&0\cr 0&1&0&0\cr 0&0&-1&0\cr
0&0&0&0\cr}\right].\cr }\eeqs  The last factor is just a gauge transformation
in $G$, so that the net effect is a change of $X$. However, there is an
ambiguity in the definition of the spin operator, since any linear combination
of $y,\tilde y$ can be added without changing its physical meaning.  A
convenient choice will be one that changes sign under parity (spin is a
pseudo--scalar). If we define the generator of rotations  to be a right
multiplication by  \beq s=\half\pmatrix{1&0&0&0\cr 0&1&0&0\cr 0&0&-1&0\cr
0&0&0&-1\cr } \eeq it will change sign under parity. (Also, left and right
multiplications by $s$ are equivalent). This matrix  differs from the naive
generator of rotations by a linear combination of $y,\tilde y$: \beq
s=\half\pmatrix{0&0&0&0\cr 0&1&0&0\cr 0&0&-1&0\cr 0&0&0&0\cr }+\half \tilde y.
\eeq

We can now solve the constraint and determine the quantum numbers of the
different states of the soliton. Any function on $G=S(U(2)\times U(2))$  can be
expanded in terms of the matrices of its irreducible  representations. Recall
that \beq (g_1,g_2,e^{i\alpha})\to \pmatrix{g_1 e^{i\alpha}&0\cr
0&g_2e^{-i\alpha}} \eeq gives a homomorphism $SU(2)\times  SU(2)\times U(1)\to
S(U(2)\times U(2))$. The kernel of this homomorphism is $(-1,-1,-1)$. So
representations of $G$ can be  labelled by $(j,j',n)$ for
$j,j'=0,\half,1\cdots$ and $n=\cdots, -1,0,1,\cdots$. In order that
$(-1,-1,-1)$ be represented by the identity,  $2(j+j')+n$ must be even. It is
useful to note that $n=2s$, $s$ being the spin variable defined earlier.

\def\mod{ \;\;{\rm mod}\;\;}

In terms of the generators of $SU(2)\times SU(2)\times U(1)$,  a basis for
wavefunctions is  labelled by $|j,j',n;j_{3L},j'_{3L};j_{3R},j'_{3R}>$. The
constraint on the wave function becomes (for $B=1$) \beq \tilde
y_R=(j_{3R}+j'_{3R})+{n\over 2}=0\quad y_R=2(j_{3R}-j'_{3R})=-k. \eeq These
conditions imply that the spin $s$ of any baryon state is integer (or
half--integer) according to whether  $k$ is integer (or half--integer). For,
the spin is $s={n\over 2}$,  which from the constraint is $(j+j')$; however,
$(j+j')=(j_{3R}+j_{3R}') \mod Z$. On the other hand
$j_{3R}+j_{3R}'=j_{3R}-j_{3R}' \mod Z$ which
in turn is ${k\over 2} \mod Z$ from
the second constraint. Thus $s={k\over 2} \mod Z$.

One representation that contains such a state is $(j,j',n)=({k\over 2},0,k)$.
Then  $j_{3R}=-{k\over 2}, j'_{3R}=0,n=k$ is the only choice that satisfies the
conditions. The different left indices will then describe a state that
transforms under the representation $({k\over 2},0,k)$ of the flavour group
$G$. This is part of a more general solution, $({k-r\over 2}, {r\over 2},k-2r)$
for $r=0,1,\cdots k$. There is exactly one state, $j_{3R}=-{k-r\over 2},
j'_{3R}={r\over 2}$ in each of these representations that satisfies the
constraint. So there is one multiplet of the left action of $G$ for each $r$.
All these representations can be grouped  into a symmetric tensor
representation of order $k$ of the group $SU(4)$ that contains $G$. These  are
therefore precisely the baryon wave functions  predicted by the naive quark
model. The soliton model predicts many more states than the naive quark model.
Only in the limit $k=N_c\to \infty$ will the two agree. This is  exactly the
analogue of what happens in the Skyrme model in $3+1$  dimensions.

The hamiltonian operator that we get from the collective action is the Laplace
operator on $G/H$. Its eigenvalues will give the mass splittings of the
different baryon multiplets.The above wavefunctions are eigenfunctions with
eigenvalues \beqs{ H|j,j',n;j_{3L},j'_{3L};j_{3R},j'_{3R}>&= \left[{1\over
2I_1}[ (j(j+1)+j'(j'+1)]+ {({n\over 2})^2\over 2I_3}\right]\cr
&|j,j',n;j_{3L},j'_{3L};j_{3R},j'_{3R}>.\cr }\eeqs The dependence of  mass
splittings  on $j,j',n$ agree with the prediction of the static quark model.
There they arise from  spin--spin coupling of the quarks due to gluon exchange.
However, the  magnitude of the splittings is much smaller in the soliton
model. In the limit that the pion mass (or current quark mass) goes to zero,
the soliton becomes very large and its moment of inertia diverges: \beq
I_{1,3}\sim {1\over F_{\pi} }\log{{m_\pi N_c}\over F_{\pi}} \eeq Then the mass
splittings go to zero as $m_{\pi}\to 0$. In the static quark model on the other
hand, even in this limit, the mass splittings are non--zero. We believe that in
$2+1$ dimensions, the soliton model is a more reliable description of the size,
and hence, the mass splittings of the baryon.

\sect { The dibaryon solution}

In the Skyrme model with $SU(2)$ symmetry, it is  known that  there is a
dibaryon solution\[braaten] with cylindrical rather than spherical symmetry.
Roughly speaking, it describes a pair of baryons rotating around each other.
The baryon number density is concentrated on a toroid. There is an analogous
dibaryon solution in 3DQCD as well.
This is a bound state of two baryons formed
by balancing their long range attraction (due to exchange of massive pions) by
a short range repulsion (due to vector meson exchange). The baryon number
density again has a maximum in a ring. However, in our case it has the same
rotational symmetry as the one baryon solution. Therefore it is a solution with
the same static ansatz as before, but with $Q=2$.
In Fig. 3 we give a comparison of the baryon number density $A'/r$ for
the baryon and dibaryon solution.
The numerical solution shows
that the baryon number density indeed has a maximum at a ring of radius about
$0.3\over m_\pi$. The dibaryon configuration has  a positive  binding energy (a
few percent  of rest mass, Fig. 4).  This shows that this is stable against
decay into a pair or baryons. For higher values of $Q$, there is a static
solution, but it is probably  unstable  with respect to non--spherically
symmetric perturbations.

The wave--functions of the dibaryon can be determined by the same collective
co--ordinate method as before. The only difference is that $N_c$ is replaced by
$2N_c$. Thus the low lying states of this dibaryon form a symmetric tensor
representation of $SU(4)$ or order $2N_c$. This is a little surprising from the
quark model point of view. The only way the quark wavefunctions can be
completely symmetric in flavor variables is for a rotational degree of freedom
to be excited. ($2N_c$ quarks cannot be put into a completely anti--symmetric
representation of color $SU(N_c)$). The natural possibility is that $N_c$ of
the quarks are in a state with orbital angular momentum $+1$ and the others in
a state with angular momentum $-1$. This agrees with the picture that the
dibaryon consists of a pair of baryons orbiting each other.

If the `baryons' can be identified as the charge carriers in a model of
superconductivity,  the dibaryon bound state we are describing can be thought
of as a Cooper pair. It will have charge $2e$ and  a  small binding energy. If
the pions are identified with the spin waves of anti--ferromagnetism, the
attractive force is ultimately anti--ferromagnetic in origin.

\sect { Soliton model with $n=1$}

The special case of $n=1$ is somewhat different because there are no
non--abelian Chern-Simons terms   that can be added to the action to recover
the correct parity properties. In this case the coset space $SU(2)/U(1)$ is
just $CP^1$, or equivalently, $S^2$. Clearly $H_4(CP^1)=0$  so that there are
no Wess--Zumino terms that can be added to the action. This case is analogous
to the Skyrme model with $SU(2)$ symmetry. As in that case, we can study this
case by embedding it into the higher flavor case. (For example, we can imagine
that one parity doublet of quarks in the case $n=2$ is much heavier than the
other.)

We can also study this case directly; $\pi_3(CP^1)=Z$ so that there is another
kind of topological  term (Hopf term) that can be added. The coefficient of
the Hopf term  can be determined by comparison to QCD.  This term  then reduces
to a model of Wilczek and Zee \[wz]. (Except that the coefficient of the Hopf
term corresponds to bosons for even $N_c$ and fermions for odd $N_c$).

The effective action then becomes  \beq S={F_{\pi}\over 2}\int \tr
\nabla_{\mu}\chi^{\dag}\nabla^{\mu}\chi d^3x + {\theta\over 4{\pi^2}} \int \bar
A d \bar A+ \half F_{\pi}m_{\pi}^2 \int \tr \chi \eps \chi^{\dag}\eps d^3x
+\cdots \eeq

Here, $\bar A={\rm Im}\; \tr \chi \pdr_{\mu}\chi^{\dag}\eps$. The second term
is the Hopf term, which is quantized. Hence $\theta$ must be periodic with
period $2\pi$. In order for parity to be a symmetry, $\theta$ must be a
multiple of $\pi$. Whether it is an even or odd multiple of $\pi$ is determined
by comparing the global anomaly of the flavor $SU(2)$ symmetry with QCD. We
will get $\theta=\pi N_c \mod 2\pi$.

At this level, this lagrangian does not contain any solitons. The Chern--Simons
terms that stabilized the soliton against collapsing to a point  have
disappeared. However at sufficiently short distances, higher derivative  terms
will become important. The first such term is a Maxwell term for the gauge
field: \beq S={F_{\pi}\over 2}\int \tr \nabla_{\mu}\chi^{\dag}\nabla^{\mu}\chi
d^3x + {\theta\over 4\pi^{2}}
\int \bar A d \bar A+ \half F_{\pi}m_{\pi}^2 \int \tr \chi \eps
\chi^{\dag}\eps d^3x -{1\over 4e^2}\int F_{\mu\nu}F^{\mu\nu} d^3x+\cdots \eeq
Lagrangians of this type have been studied in the  literature in a cosmological
context \[vachaspati].  Except for the Hopf term, our effective lagrangian
describes an abelian Higgs model (with an infinite mass for the Higgs field)
with a global $SU(2)$ symmetry, broken by the pion mass. There are solitons in
this theory of mass about $  F_{\pi}(a +b { m_\pi\over e})$, $a$ and $b$ being
constants of order one.  If the $SU(2)$ symmetry is exact, (the  pion mass is
zero)  the soliton is infinitely large. (These are called `semi--local
strings' in the cosmological context).  Otherwise they will have a size of
order ${1\over m_{\pi}}$. The same arguments that we made earlier will show
that these are fermions or bosons depending on whether $N_c$ is even or odd.
The  collective variable $X$ is valued in $U(1)\times U(1)$; the wave functions
of the solitons will be spanned by the basis $|j_3,j'_3>$ for
$j_3,j'_3=0,\half,{3\over 2}\cdots$. However, the Hopf term will require that
$j_3+j_3'={N_c\over 2} \mod Z$. Thus we have integer spin for even $N_c$ and
half--integer spin for odd $N_c$.

\sect { Quark Model of Baryons}

The naive quark model of baryons assumes that the flavor symmetry breaking
results in constituent masses to the quarks. Massive quarks carry spin. In
order to preserve parity, we assume  $n$ quarks are spin up and the other $n$
are spin down. The flavor symmetry is now $U(n)\times U(n)= U(1)\times G$. The
first $U(1)$ is baryon number symmetry. The $U(1)$ factor within $G$ is
interpreted as spin.

We first discuss the constituent quark mass. In 3+1 dimensions, the constituent
quark mass $M$ is roughly $M+ m= M_{B}/N_{c}$, where $M_{B}$ is the mass of
some baryon and $m$ is the current quark mass. It agrees with  the Skyrme
model. In our case, the soliton model predicts some thing different:
$M_{B}/N_{c} \sim \tilde{F}_{\pi} +  \sqrt (m\tilde{F}_{\pi})$. (In this
section we denote $F_\pi/N_c$ by $\tilde F_\pi$; this is the quantity that is
kept finite in the large $N_c$ limit.) Thus the
dependence of the constituent quark mass on the current quark mass has an
exponent $\half$ instead of $1$ as one would naively expect. This is
reminiscent of anomalous critical exponents in the theory of phase transitions.

The baryon wave functions from the nonrelativistic quark model agree with the
low lying modes of soliton model. As in 3+1 dimensions, we assume that baryons
form completely symmetric representation under $G$ since the quarks are
antisymmetric in color. It is more convenient to look for the symmetric
representation of $SU(2n)$ which combines the flavor and spin symmetries. The
Young tableau consists $N_{c}$ boxes in the same row. Each box represents
$2n$-dimensional representation of $SU(2n)$, corresponding to $2n$ quark
states. We can then decompose it to irreducible representations  of $G$.
Consider $n=2$, $G=S(U(2)\times U(2))$. We can label the $SU(4)$ symmetric
representation with $N_{c}$ boxes by a sum of $|j, j^{\prime}, n>$, where $j,
j^{\prime}$ denote spin $j$ and $j^{\prime}$ representations  of the two
$SU(2)$, respectively, and $n$ labels the remaining $U(1)$ group. Since there
are $N_{c}$ boxes, \beq j+j^{\prime}={N_{c}\over 2}. \eeq The spin is \beq s=
{n \over 2} = j-j^{\prime}. \eeq These agree with the solutions to the
constraints found in the soliton model.  Of course, the soliton model  has more
states and there is complete agreement only in the limit $N_c\to \infty$.

One of the interesting predictions of the naive quark model in 3+1 dimensions
is the spin-dependent mass splitting. It comes from the residual gluon
exchange. If we simply take over it to 2+1 dimensions, the Hamiltonian is \beq
{\mu\over M^{2}} \sum s_{i}s_{j} \sim {\mu\over M^{2}}  S^{2} + const., \eeq
where $\mu$ is some constant of dimension 3, $S$ is the spin of the baryon.
This is, however, different from the soliton prediction. The moment of inertia
is  $ \ln (\tilde{F}_{\pi}/m_{\pi})/\tilde{F}_{\pi}$ there. The soliton model
predicts a spin--splitting that is smaller by a factor  $\ln{\tilde
F_{\pi}\over m_{\pi}}$.

One can also work out the prediction of the  magnetic  moments. Let us consider
electromagnetic $U(1)$ charge  to be  the baryon number. (For simplicity we
consider only the iso--singlet part of electric charge). In the quark model,
one simply writes the interacting Hamiltonian as \beq H_{int} =\sum {s_{i}\over
M}B = B{S\over M}. \eeq In the soliton model, the $U(1)$ current is just the
topological current (19). The gauge coupling is then a Chern--Simons term, \beq
S_{int}={1\over 2\pi}\int
d^{3}x\epsilon^{\mu\nu\rho}A^{em}_{\mu}\partial_{\nu}A_{3}^{\rho}. \eeq  Since
$A_3$ is pseudo-vector, this  is parity invariant. The interaction is gauge
invariant provided $\partial_{\nu}A_{3}^{\rho}$ is well-behaved at infinity. We
are interested in $B_{i}$ non-zero, i.e., magnetic coupling. We have to compute
the current $j_{i}$. For $\chi=X(t)\chi_{0}$, the equation of motion tells us
\beq A_{3i}=\tr \chi^{\dagger}_{0}\partial_{i}\chi_{0}, A_{30}=\tr
\chi_{0}^{\dagger} X^{\dagger}\dot{X}\chi_{0}\epsilon. \eeq Since $A_{3i}$ is
time independent, only $A_{30}$ contributes to the current. One can show that
\beq j_{i}=\partial_{i}A_{30}=(\sin^{2}\phi)^{\prime}\tr
X^{\dagger}\dot{X}\pmatrix{0&0&0&0\cr 0&1&0&0\cr 0&0&-1&0\cr 0&0&0&0\cr}. \eeq
The interaction Lagrangian is \beq S_{int}={1\over 2\pi}\int dt d^{2}x
A_{i}^{em}\epsilon^{ij}\partial_{j} (\sin^{2}\phi) \tr X^{\dagger}\dot{X}
\pmatrix{0&0&0&0\cr 0&1&0&0\cr 0&0&-1&0\cr 0&0&0&0\cr}. \eeq One can integrate
by parts, since $\sin \phi$ is well behaved when $m_{\pi}$ is not zero. For
constant magnetic field, the result is \beq S_{int} \sim \int dt
{\ln({\tilde{F}_{\pi}\over m_{\pi}})\over \tilde{F}_{\pi}^{2}} B \tr
X^{\dagger}\dot{X} \pmatrix{0&0&0&0\cr 0&1&0&0\cr 0&0&-1&0\cr 0&0&0&0\cr}. \eeq
Recall the definition of the  angular momentum $S$, we have the interacting
Hamiltonian \beq H_{int} \sim B S/\tilde{F}_{\pi}, \eeq which qualitatively
agrees with the quark model prediction. Thus the soliton model predicts that
the  color and magnetic moment of the constituent quark is smaller by a factor
of $\log {\tilde F_{\pi}\over m_{\pi}}$ than one would naively expect.

At long distances, the force between two constituent quarks has an additional
piece mediated by pion exchange. If it is attractive, one has an explanation of
how  dibaryon bound state is formed (equivalently one can check the long range
force between two baryons). For simplicity we consider $N=2$ and $N_{c}=3$
case. There are two pions $\pi^{\pm}$. We write down the effective Lagrangian
of the pions and the constituent quarks $\psi_{u}$ and  $\psi_{d}$, \beq S=
\int d^{3}x (\bar{\Psi}\gamma^{\mu}\partial_{\mu}\Psi + M\bar{\Psi} {\bf
\pi}\cdot {\bf\tau} \Psi + F_{\pi} \partial_{\mu}{\bf \pi}\cdot
\partial_{\mu}{\bf \pi} + F_{\pi}m^{2}_{\pi}\pi^{+}\pi^{-}), \eeq where \beq
\Psi=\pmatrix{\psi_{u}\cr \psi_{d}}, {\bf \pi}=(\pi^{+},\pi^{-},\pi_{3}),
\pi_{3}^{2}+\pi^{+}\pi^{-}=1, \eeq and $\tau$'s are Pauli matrices. We have
omitted residual color force. One can easily compute the force between the two
quarks. We first do non-relativistic reduction of $\Psi$, then integrate over
$\pi$'s, it turns out that the only force is an attractive one between u-quark
and d-quark, \beq U(r) \sim -{m_{\pi}^{2}\over F_{\pi}} e^{-m_{\pi}r}. \eeq
This attractive force provides a mechanism for the formation of dibaryons.

In summary, the soliton model predicts a larger contribution  of current quark
mass to the baryon mass, and moment of inertia is much larger than the naive
quark model. The soliton model arises in a systematic $1/N_{c}$ expansion,
whereas  the naive quark model doesn't seem to become exact in any limit. We
can  conclude that in 2+1 dimensions (or lower), soliton model describes
baryons more accurately than the naive quark model, or its refinement,  the
analog of chiral quark model in 3+1 dimensions. The underlying reason for this
is the infrared behavior of pions: the pion condensate in $d\ge 2+1$ is
negligible at long distance, while it is more important in $d\le 2+1$. $d=2+1$
is in some sense a critical dimension, with a logarithmic divergence in the
moments of inertia.

\sect{ Acknowledgements}

S.G.R. thanks P. Wiegmann for rekindling his interest in soliton models. We
also
thank A.P. Balachandran  and K. Gupta for discussions.
This work is supported in part by the US Department of Energy Contract
No. DE-AC02-76ER13065.

\vfill\eject

\centerline{\bf Figure Captions} \vskip0.5in

{\bf Fig. 1} One baryon
solution for $2\pi F_\pi /N_c m_\pi = 7$. The solid line represents the
function
$\phi$ and the dashed line represent the function $A$. \vskip0.5in

{\bf Fig. 2} Baryon energy in units of $m_\pi$ as function of the ratio
$2\pi F_\pi /N_c m_\pi$. \vskip0.5in

{\bf Fig. 3} Baryon number density for the single baryon (solid line) and
dibaryon (dashed line) solutions. The choice of parameters is as before: $2\pi
F_\pi /N_c m_\pi = 7$. \vskip0.5in

{\bf Fig. 4} Binding energy of the dibaryon as function of the ratio
$2\pi F_\pi /N_c m_\pi$. Here, $\Delta E= 2E_{Q=1} - E_{Q=2}$

\vfill\eject

{\it References}\hfill\break

\refis{I} G. Ferretti, S.G.Rajeev and Z. Yang, {\it Effective Lagrangian for
Three Dimensional Quantum Chromodynamics}, U. of R. Preprint April 1992.

\refis{balreview} A.P.Balachandran, Tasi   lectures, ed. F. Gursey and M.
Bowick
(World Scientific, Singapore, 1985).

\refis{hongetal} J. Hong, Y. Kim and P.Y. Pac, Phys. Rev. Lett. 64 (1990) 2230.

\refis{jackiwetal} R. Jackiw and E. Weinberg, Phys. Rev. Lett. 64 (1990) 2234.

\refis{goldwilcz} J. Goldstone and F. Wilczek, Phys. Rev. Lett. 47 (1981) 986.

\refis{baletal} A. P. Balachandran, V.P. Nair, S. G. Rajeev and A. Stern, Phys.
Rev. Lett. 49 (1982) 1124; Phys. Rev. D27 (1983) 1153.

\refis{witten} E. Witten, Nucl. Phys. B223 (1983) 422, 433.

\refis{Al} A. Stern, Phys. Rev. Lett. 59 (1987) 1506.

\refis{spinstatistics}  A.P. Balachandran, Int.J.Mod.Phys.B5 (1991) 2585.

\refis{rajaraman} R. Rajaraman,  Solitons and Instantons (North-Holland,
Amsterdam, 1982).

\refis{perelemov} A.M. Perelemov, Phys. Rep. 146 (1987) 135.

\refis{braaten} E. Braaten and L. Carson, Phys. Rev. D37 (1988) 3349.

\refis{vachaspati} T. Vachaspati and A. Achucarro, Phys. Rev. D44 (1991) 3067.

\refis{wz} F. Wilczek and A. Zee, Phys. Rev. Lett. 51 (1983) 2250.

\bye